\def\year{2022}\relax
\documentclass[letterpaper]{article} 
\usepackage{aaai22}  
\usepackage{times}  
\usepackage{helvet}  
\usepackage{courier}  
\usepackage[hyphens]{url}  
\usepackage{graphicx} 
\usepackage{natbib}  
\usepackage{caption} 
\DeclareCaptionStyle{ruled}{labelfont=normalfont,labelsep=colon,strut=off} 
\usepackage[ruled,linesnumbered,commentsnumbered]{algorithm2e}

\usepackage{newfloat}
\usepackage{listings}
\usepackage{graphicx}
\usepackage{amssymb}
\usepackage{booktabs}
\usepackage{multirow,array}
\usepackage{mathtools}
\usepackage{array,ragged2e}
\usepackage{tablefootnote}
\usepackage{url}
\usepackage{float}
\usepackage{graphicx}
\usepackage{times}
\usepackage{helvet}
\usepackage{courier}
\usepackage{tcolorbox}
\usepackage{amsmath}
\usepackage{bbding}
\usepackage{balance}

\DeclareRobustCommand\onedot{\futurelet\@let@token\@onedot}
 
\def\ie{\emph{i.e.}}

\floatstyle{ruled}
\newfloat{listing}{tb}{lst}{}
\floatname{listing}{Listing}
\usepackage[ruled,linesnumbered,commentsnumbered]{algorithm2e}

\title{Heterogeneity-aware Twitter Bot Detection with Relational Graph Transformers}
\author {
Shangbin Feng,\textsuperscript{\rm 1}
Zhaoxuan Tan,\textsuperscript{\rm 1}
Rui Li,\textsuperscript{\rm 2}
Minnan Luo\textsuperscript{\rm 1}
}
\affiliations{
    \textsuperscript{\rm 1}School of Computer Science and Technology, Xi'an Jiaotong University, Xi'an, China\\
    \textsuperscript{\rm 2}School of Continuing Education, Xi'an Jiaotong University, Xi'an, China\\

    \{wind$\_$binteng, tanzhaoxuan\}@stu.xjtu.edu.cn, \{lrvberg,minnluo\}@xjtu.edu.cn
}

\usepackage{bibentry}

\begin{document}

\maketitle

\begin{abstract}
Twitter bot detection has become an important and challenging task to combat misinformation and protect the integrity of the online discourse. 
State-of-the-art approaches generally leverage the topological structure of the Twittersphere, while they neglect the heterogeneity of relations and influence among users.
In this paper, we propose a novel bot detection framework to alleviate this problem, which leverages the topological structure of user-formed heterogeneous graphs and models varying influence intensity between users.
Specifically, we construct a heterogeneous information network with users as nodes and diversified relations as edges.
We then propose relational graph transformers to model heterogeneous influence between users and learn node representations.
Finally, we use semantic attention networks to aggregate messages across users and relations and conduct heterogeneity-aware Twitter bot detection.
Extensive experiments demonstrate that our proposal outperforms state-of-the-art methods on a comprehensive Twitter bot detection benchmark.
Additional studies also bear out the effectiveness of our proposed relational graph transformers, semantic attention networks and the graph-based approach in general.
\end{abstract}

\section{Introduction}
Twittier bots are Twitter accounts controlled by automated programs or the Twitter API. Bot operators often launch bot campaigns to pursue malicious goals, which harms the integrity of the online discourse. Over the past decade, Twitter bots were actively involved in election interference~\cite{10.1145/3308560.3316486,DBLP:journals/corr/Ferrara17aa}, spreading misinformation~\cite{10.1145/3409116} and promoting extreme ideology~\cite{berger2015isis}. Since malicious Twitter bots pose threat to online communities and induce undesirable social effects, effective Twitter bot detection measures are desperately needed.

Earlier works in Twitter bot detection generally rely on feature engineering, where an ample amount of user features are proposed and evaluated. Features extracted from tweets~\cite{cresci2016dna} and user metadata~\cite{yang2020scalable,lee2013warningbird,miller2014twitter} were combined with traditional classifiers for bot detection. With the advent of deep learning, neural network based Twitter bot detectors were increasingly prevalent. Recurrent neural networks are adopted to encode tweets and detect bots based on their semantic content~\cite{kudugunta2018deep,wei2019twitter}. Self-supervised learning techniques were introduced to counter bot evolution~\cite{feng2021satar}. 
Graph neural networks~\cite{ali2019detect,feng2021botrgcn} were later used to leverage the graph structure of the Twittersphere, while state-of-the-art methods are topology-aware in one way or another.

\begin{figure}[t]
    \centering
    \includegraphics[width = 1\linewidth]{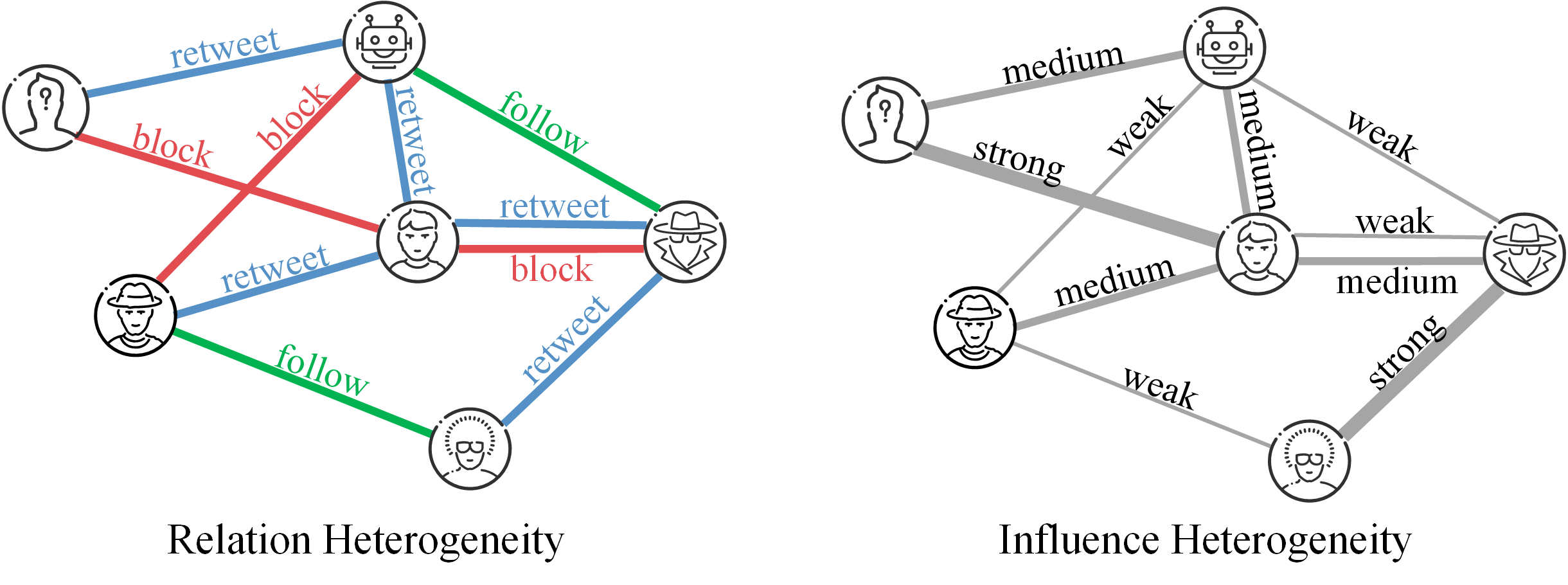}
    \caption{Users and bots of real-world social media interact in different ways and have varied influence over others, which result in relation and influence heterogeneity.}
    \label{heterogeneity_aware}
\end{figure}

Despite earlier successes of leveraging the topological structure of the Twittersphere, these methods fail to recognize the intrinsic heterogeneity of Twitter and leverage it to identify subtle differences between genuine users and novel Twitter bots. Figure \ref{heterogeneity_aware} illustrates two levels of heterogeneity that are pervasive on the real-world Twittersphere:
\begin{itemize}
    \item \textbf{Relation Heterogeneity.} Twitter users are connected with different types of relations. For example, one user might like, comment, retweet or block another user, while these activities signal different relations between them.
    \item \textbf{Influence Heterogeneity.} Twitter users have different influence range and intensity over their neighbors on the Twittersphere. For example, distinguished news outlets might have a tremendous impact on the minds of many, while ordinary users generally inform close circles of their recent activities.
\end{itemize}

\begin{figure*}[t]
    \centering
    \includegraphics[width=0.8\linewidth]{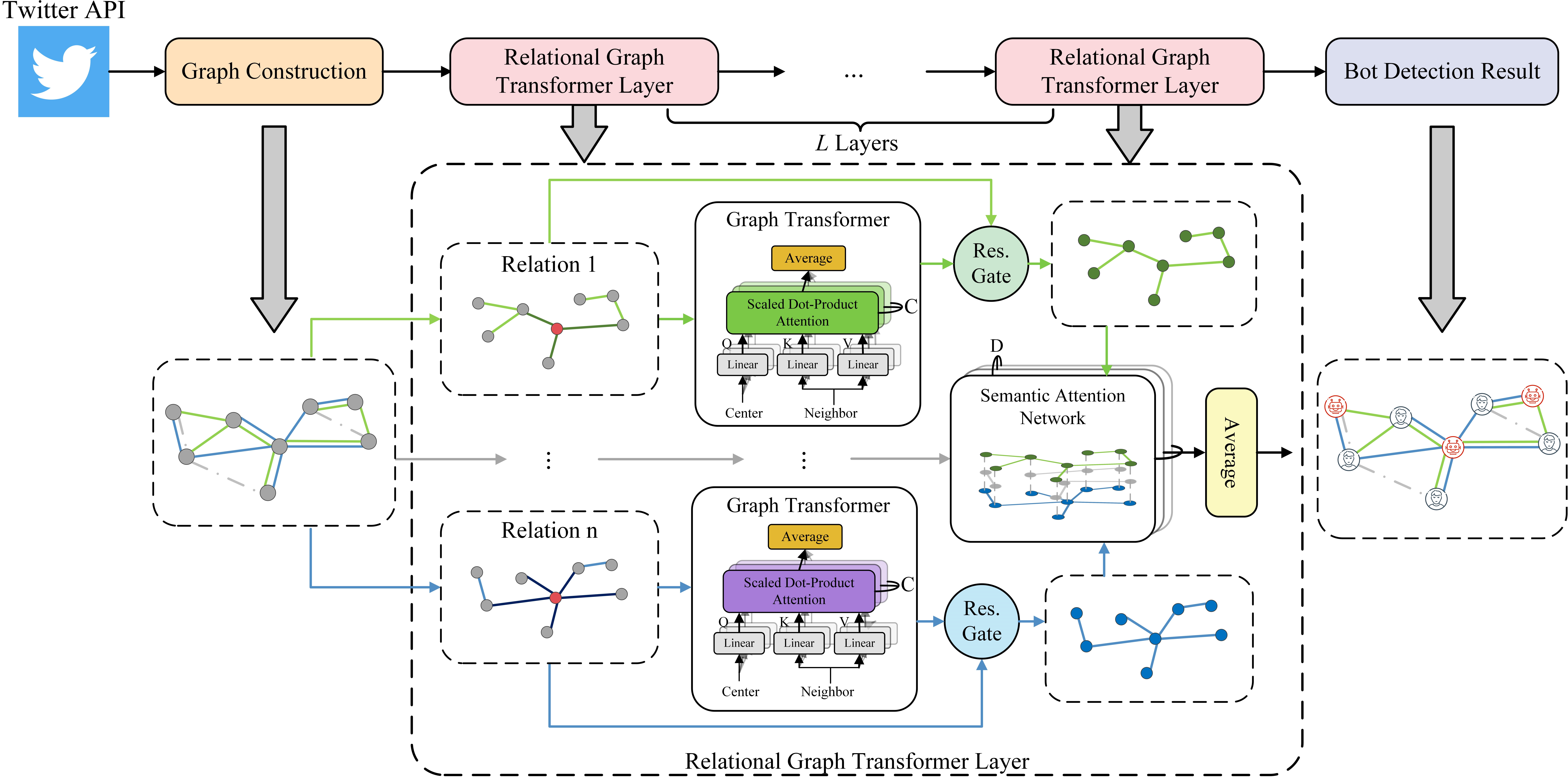}
    \caption{Overview of our graph-based and heterogeneity-aware Twitter bot detection framework.}
    \label{model_overview}
\end{figure*}

In this paper, we propose a novel Twitter bot detection framework that leverages the topological structure of the real-world Twittersphere, and on top of that, models pervasive heterogeneity of relation and influence to boost task performance. Specifically, we construct heterogeneous information networks with users as nodes and diversified relations as edges. We then propose relational graph transformers to model influence intensity with the attention mechanism and learn node representations. Finally, we adopt semantic attention networks to aggregate messages across users and relations and conduct bot detection. Our main contributions are summarized as follows:
\begin{itemize}
    \item We propose to leverage relation and influence heterogeneity of the real-world Twittersphere, which enables our bot detection model to identify subtle differences between genuine users and bots and conduct robust bot detection.
    \item We propose a novel Twitter bot detection framework that is graph-based and heterogeneity-aware. It is an end-to-end bot detector that adopts relational graph transformers to leverage the topology and heterogeneity of the real-world Twittersphere.
    \item We conduct extensive experiments to evaluate our model and state-of-the-art methods on a comprehensive bot detection benchmark. Results demonstrate that our proposal consistently outperform all baseline methods. Further experiments also bear out the effectiveness of our graph-based and heterogeneity-aware approach.
\end{itemize}

\section{Related Work}

\subsection{Twitter Bot Detection}
Early Twitter bot detection models focus on manually designed features and combine them with traditional classifiers. These features are extracted from tweets~\cite{cresci2016dna}, user metadata~\cite{yang2020scalable,lee2013warningbird} or both~\cite{miller2014twitter}. As deep learning later shows great promise and gains popularity, an increasing amount of neural network based bot detectors are proposed. Fully connected networks~\cite{kudugunta2018deep}, recurrent neural networks~\cite{wei2019twitter} and generative adversarial networks~\cite{stanton2019gans} are adopted in effective bot detection models to leverage different aspects of user information. SATAR~\cite{feng2021satar}, a recently proposed framework, jointly leverages multi-modal user information with different deep architectures to improve upon these methods.

Although SATAR~\cite{feng2021satar} proposes to leverage the graph structure of the Twittersphere for bot detection, it does so in a feature engineering manner, rather than adopting state-of-the-art graph neural network architectures. Graph-based bot detectors were proposed to fill in the blanks. \cite{ali2019detect} views Twitter as a network of users and adopt graph convolutional networks to conduct bot detection. \cite{feng2021botrgcn} further constructs a heterogeneous information network to represent Twitter and uses relational GNNs for bot detection, which achieves state-of-the-art performance. However, these graph-based methods fail to incorporate the intrinsic heterogeneity of relation and influence on the real-world Twittersphere. In this paper, we build on these works and propose a heterogeneity-aware bot detector, which dynamically incorporates and leverages diversified relations and influence patterns between users.

\subsection{Heterogeneous Information Networks}
Real-world network data often consist of large quantities of diversified and interactive entities, which can be called heterogeneous information networks (HINs). HINs are widely adopted to model social networks~\cite{survey6,survey7,nguyen2020fang}, link and graph mining~\cite{survey3,survey10} and natural language processing systems~\cite{de2018question,feng2021knowledge}. To effectively analyze HINs, ~\cite{RGCN} proposes relational graph convolutional networks to extend GCN~\cite{GCN} to heterogeneous graphs. ~\cite{wang2019heterogeneous} proposes heterogeneous graph attention networks to extend GAT to heterogeneous graphs. In this paper, we build on these works to propose relational graph transformers and leverage Twitter heterogeneity.


\section{Methodology}

\subsection{Overview}
Figure \ref{model_overview} presents an overview of our proposed graph-based and heterogeneity-aware Twitter bot detector. Specifically, we firstly construct a heterogeneous information network with diversified relations to represent the Twittersphere. We then learn node representations under each relation with our proposed relational graph transformers. After that, we take a global view of the graph and dynamically aggregates representations across relations with semantic attention networks. Finally, we classify Twitter users into bots or genuine users and learn model parameters.

\subsection{Graph Construction}
We construct a heterogeneous information network (HIN) to represent the Twittersphere, which takes the relation heterogeneity into account and leverages diversified interactions between users. Specifically, we take Twitter users as nodes in the graph and we connect them with different types of edges, representing diversified relations on Twitter. We denote the set of relations in the HIN as $R$ while our framework supports any relation settings.

Since this paper focuses on leveraging relation and influence heterogeneity to improve bot detection, we follow the same user information encoding procedure in the state-of-the-art approach~\cite{feng2021botrgcn} for fairness. We denote user $i$'s feature vector as $x_i$ and transform it with a fully connected layer to serve as initial features in the GNNs, \ie,
\begin{equation}
    x_i^{(0)}=\sigma(W_{I}\cdot x_i + b_{I})
    \label{in_linear}
\end{equation}

\noindent where $W_I$ and $b_I$ are learnable parameters, $\sigma$ denotes non-linearity and we use leaky-relu as $\sigma$ without further notice.

\subsection{Relational Graph Transformers}
Inspired by Transformers~\cite{vaswani2017attention} and its success in natural language processing, we propose relational graph transformers, a GNN architecture that incorporates transformers and operates on HINs. We firstly obtain query, key and value for the $c$-th attention head with regard to relation $r$ and node $i$, formulated as
\begin{align}
\begin{aligned}
    {q_{c,i}^{r}}^{(l)}&={W_{c,q}^{r}}^{(l)}\cdot x_i^{(l-1)} +{b_{c,q}^{r}}^{(l)},\\
    {k_{c,j}^{r}}^{(l)}&={W_{c,k}^r}^{(l)}\cdot x_j^{(l-1)}+{b_{c,k}^r}^{(l)},\\
    {v_{c,j}^{r}}^{(l)}&={W_{c,v}^r}^{(l)}\cdot x_{j}^{(l-1)}+{b_{c,v}^{r}}^{(l)},
\end{aligned}
\label{kqv}
\end{align}
where $q$, $k$ and $v$ are query, key and value of the attention mechanism, $(l)$ denotes the $l$-th layer of GNNs, all $W$ and $b$ are learnable parameters with regard to different relations and attention heads. 
We then model influence heterogeneity by calculating attention weights between different nodes by 
\begin{equation}
    {\alpha_{c,ij}^{r}}^{(l)}=\frac{\langle {q_{c,i}^{r}}^{(l)}, {k_{c,j}^{r}}^{(l)}\rangle}{\sum_{u\in N^{r}(i)} \langle {q_{c,i}^r}^{(l)}, {k_{c,u}^{r}}^{(l)} \rangle},
    \label{coeff}
\end{equation}
where ${\alpha_{c,ij}^{r}}^{(l)}$ denotes the attention weight between nodes $i$ and $j$, $\langle q,k\rangle = \exp(\frac{q^T k}{\sqrt{d}})$ is the exponential scale dot-product function where $d$ is the hidden size of each attention head, $N^r(i)$ denotes node $i$'s neighborhood with regard to relation $r$.  
We then aggregate over node neighborhood and attention heads to obtain node representation under relation $r$, \ie,
\begin{equation}
    {u_{i}^{r}}^{(l)}=\frac{1}{C}\sum_{c=1}^{C}\left[ \sum_{j\in N^r(i)} {\alpha_{c,ij}^{r}}^{(l)} \cdot {v_{c,j}^{r}}^{(l)}\right],
    \label{transout}
\end{equation}
where $ {u_{i}^{r}}^{(l)}$ is the hidden representations of node $i$ in $l$-th layer for relation $r$, $C$ is the number of attention heads. We then apply the gate mechanism to obtained results to ensure smooth representation learning. We firstly obtain the gate level as follows
\begin{equation}
    {z_{i}^{r}}^{(l)}=sigmoid (W_{A}^{r}\cdot [{u_{i}^{r}}^{(l)}, {x_i^{(l)}}]+b_{A}^{r}),
    \label{gate1}
\end{equation}
where $[\cdot,\cdot]$ is the concatenation operation, $W_A$ and $b_A$ are learnable parameters. We then apply the gate mechanism to learned representation ${u_i^r}^{(l)}$ and input ${x_i^r}^{(l)}$ by
\begin{equation}
    {h_i^r}^{(l)}=tanh ({u_i^r}^{(l)})\odot {z_{i}^{r}}^{(l)}+{x_{i}^{r}}^{(l)}\odot (1-{z_{i}^{r}}^{(l)}),
    \label{gate2}
\end{equation}
where $\odot$ denotes the Hadamard product operation and ${h_i^r}^{(l)}$ is the learned representation of node $i$ with regard to relation $r$ in the $l$-th layer.

\begin{algorithm}[!t]
\caption{Model Learning Algorithm}
\label{algo}
\SetKwInOut{Input}{input}\SetKwInOut{Output}{output}
\SetAlgoLined

\Input{Twitter bot detection dataset $T$}
\Output{Optimized model parameters $\theta$}

initialize $\theta$; \\
construct Twitter HIN to obtain relations $R$;\\ 
encode user information to obtain $x_i$;\\
$x_i^{(0)}\gets$Equation (\ref{in_linear});\\
\While{$\theta$ has not converged}{
\For{$r\in R$}{
\For{each user $i\in T$}{
find relation-based neighborhood $N^r(i)$;\\
\For{$c\gets 1$ \KwTo $C$}{

\For{$j \in N^r(i)$}{
\mbox{
${q_{c,i}^{r}}^{(l)}$, ${k_{c,j}^{r}}^{(l)}$, ${v_{c,j}^{r}}^{(l)}\gets$  Equation  (\ref{kqv});}\\

${\alpha_{c,ij}^{r}}^{(l)}\gets$Equation (\ref{coeff});\\
}
}
${h_i^r}^{(l)}\gets$ Equation (\ref{transout} - \ref{gate2});
}

\For{$d\gets 1$ \KwTo $D$}{
$\beta_{d}^{r}\gets$Equation (\ref{weight} - \ref{beta});\\
}
}
$x_i^{(L)}\gets$ Equation (\ref{sem});\\


$Loss\gets$ Equation (\ref{softmax} - \ref{loss});\\
$\theta\gets$ BackPropagate ($Loss$);\\

}
\Return $\theta$
\end{algorithm}

\subsection{Semantic Attention Networks}
After analyzing the HIN while separating different relations, we use semantic attention networks to aggregate node representations across relations while preserving the relation heterogeneity entailed in the Twitter HIN. Firstly, we obtain the importance of each relation by taking a global view of all nodes in the HIN, \ie,
\begin{equation}
    {w_{d}^r}^{(l)}=\frac{1}{\vert V \vert}\sum_{i\in V} {q_{d}^{(l)}}^{T}\cdot tanh ({W_{d,s}}^{(l)}\cdot{h_{i}^{r}}^{(l)}+{b_{d,s}}^{(l)}),
    \label{weight}
\end{equation}
where ${w_{d}^r}^{(l)}$ denotes the weight of relation $r$ at the $d$-th attention head, $V$ denotes the set of nodes in HIN, ${q_d^{(l)}}$ is the semantic attention vector at the $d$-th attention head in layer $l$, $q_d^{(l)}$, $W_{d,s}^{(l)}$ and $b_{d,s}^{(l)}$ are learnable parameters of the semantic attention network. We normalize the importance of each relation with softmax, formulated by
\begin{equation}
    {\beta_{d}^{r}}^{(l)} = \frac{\exp({w_{d}^{r}}^{(l)})}{\sum_{k 
    \in R}\exp({w_{d}^{k}}^{(l)})},
    \label{beta}
\end{equation}
where ${\beta_{d}^{r}}^{(l)}$ denotes the weight of relation $r$. We then fuse node representations under different relations with these weights as follows
\begin{equation}
    x_i^{(l)}=\frac{1}{D}\sum_{d=1}^{D}\left[\sum_{r \in R}{\beta_{d}^{r}}^{(l)}\cdot {h_i^r}^{(l)}\right],
    \label{sem}
\end{equation}
where $x_i^{(l)}$ denotes the output of layer $l$, ${h_i^r}^{(l)}$ denotes the results of relational graph transformers and $D$ is the number of attention heads in the semantic attention network.

\subsection{Learning and Optimization}
Each layer of GNN in our model contains a relational graph transformer and a semantic attention network. After $L$ layers of GNNs, we obtain the final node representations $x^{(L)}$. We transform them with an output layer and a softmax layer for Twitter bot detection, \ie,
\begin{equation}
    \hat{y_i}=softmax(W_{O}\cdot \sigma(W_{L}\cdot x_{i}^{(L)}+b_{L}) + b_{O}),
    \label{softmax}
\end{equation}
where $\hat{y_i}$ is our model's prediction of user $i$, all $W$ and $b$ are learnable parameters. We then train our bot detector with supervised annotations and a regularization term, formulated as
\begin{equation}
    Loss = -\sum_{i\in Y}\left[ y_i\log(\hat{y_i})+(1-y_i)\log(1-\hat{y_i})\right]+\lambda\sum_{w\in \theta}w^2,
    \label{loss}
\end{equation}
where $Y$ is the annotated user set, $y_i$ is the ground-truth labels, $\theta$ denotes all trainable parameters in the model and $\lambda$ is a hyperparameter. To sum up, Algorithm \ref{algo} presents the overall training schema of our proposed graph-based and heterogeneity-aware bot detection framework, with time complexity of $O(\vert E\vert)$ for each layer where $E$ denotes the edge set, assuming embedding dimension and the number of relations are constants.

\section{Experiments}

\subsection{Dataset}
Our bot detection model is graph-based and heterogeneity-aware, which requires data sets that provide certain type of graph structure. TwiBot-20~\cite{feng2021twibot} is a comprehensive Twitter bot detection benchmark and the only publicly available bot detection dataset to provide user follow relationships to support graph-based methods. In this paper, we make use of TwiBot-20, which includes 229,573 Twitter users, 33,488,192 tweets, 8,723,736 user property items and 455,958 follow relationships. We follow the same splits provided in the benchmark so that results are directly comparable with previous works.

\begin{table*}[t]
    \centering
	\caption{Characteristic and performance of different Twitter bot detection methods. Deep, interactive, representative, graph-based and heterogeneity-aware denotes whether the method involves deep learning, leverages user interactions, learns user representation, involves graph neural networks or leverages Twitter heterogeneity.}
	\begin{tabular}{lccccccc}
		\toprule[1.5pt]
		\textbf{Method} & \textbf{Deep} & \textbf{Interactive} & \textbf{Representative} & \textbf{Graph-based} & \textbf{Heterogeneity-aware}&\textbf{Accuracy} & \textbf{F1-score} \\%
		\midrule[0.75pt]
		Lee \textit{et al.}& & & & & & 0.7456 & 0.7823\\
		Yang \textit{et al.}& & & & & & 0.8191 & 0.8546\\
		Cresci \textit{et al.}& & & & & & 0.4793 & 0.1072\\
		Kudugunta \textit{et al.}&\checkmark & & & & & 0.8174 & 0.7515\\
		Wei \textit{et al.}&\checkmark & & & & & 0.7126 & 0.7533\\
		Miller \textit{et al.}& & \checkmark& & & & 0.4801 & 0.6266\\
		Botometer & &\checkmark & & & & 0.5584 & 0.4892\\
		SATAR &\checkmark & \checkmark& \checkmark& & & 0.8412 & 0.8642\\
		Alhosseini \textit{et al.}& \checkmark&\checkmark &\checkmark & \checkmark& & 0.6813 & 0.7318\\
		BotRGCN &\checkmark &\checkmark &\checkmark &\checkmark & & 0.8462 & 0.8707\\
		\midrule[0.75pt]
		\textbf{Ours}&\checkmark &\checkmark & \checkmark &\checkmark & \checkmark & \textbf{0.8664} & \textbf{0.8821}\\
		\bottomrule[1.5pt]
		\label{bigtable}
	\end{tabular}
\end{table*}

\begin{table}[t]
	\centering
	\caption{Hyperparameter settings of our model. We make use of follower and following information as relations $R$, while we discuss more choices in heterogeneity study.}
	\begin{tabular}{c|c}
		\toprule[1.5pt]
		\textbf{Hyperparameter} & \textbf{Value}\\
		\midrule[0.75pt]
		optimizer & AdamW\\
		learning rate & $\rm{10^{-3}}$\\
		$L_2$ regularization $\lambda$ & $\rm{3\times 10^{-5}}$\\
		batch size & 256\\
		layer count $L$ & 2\\
		dropout & 0.5\\
		size of hidden state & 128\\
		maximum epochs & 40\\
		transformer attention heads $C$ &8\\
		semantic attention heads $D$&8\\
		relational edge set $R$ & $\{follower, following\}$\\
		\bottomrule[1.5pt]
	\end{tabular}
	\label{implement}
\end{table}

\subsection{Baselines}
We compare our graph-based and heterogeneity-aware approach with the following methods:

\begin{itemize}
\item \textbf{Lee \textit{et al.}}~\cite{lee2011seven} extract features from Twitter user such as the longevity of account and combine them with random forest classifier.

\item \textbf{Yang \textit{et al.}}~\cite{yang2020scalable} use random forest classifier with minimal user metadata and derived features.

\item \textbf{Cresci \textit{et al.}}~\cite{cresci2016dna} encode user activity sequences with strings and identify longest common substrings to identify bot groups.

\item \textbf{Kudugunta \textit{et al.}}~\cite{kudugunta2018deep} propose to jointly leverage user tweet semantics and user metadata.

\item \textbf{Wei \textit{et al.}}~\cite{wei2019twitter} use recurrent neural networks to encode tweets and classify users based on their tweets.

\item \textbf{Miller \textit{et al.}}~\cite{miller2014twitter} extract 107 features from user tweets and metadata and frames the task of bot detection as anomaly detection.

\item \textbf{Botometer}~\cite{davis2016botornot} is a bot detection service that leverages more than 1,000 user features.

\item \textbf{SATAR}~\cite{feng2021satar} is a self-supervised representation learning framework of Twitter users that jointly leverages user tweets, metadata and neighborhood information. SATAR conducts bot detection by fine-tuning on specific bot detection data sets.

\item \textbf{Alhosseini \textit{et al.}}~\cite{ali2019detect} use graph convolutional networks to learn user representations and conduct bot detection.

\item \textbf{BotRGCN}~\cite{feng2021botrgcn} constructs a heterogeneous graph to represent the Twittersphere and adopts relational graph convolutional networks for representation learning and bot detection. BotRGCN achieves state-of-the-art performance on the comprehensive TwiBot-20 benchmark.
\end{itemize}

\subsection{Implementation}
We use pytorch~\cite{paszke2019pytorch}, pytorch lightning~\cite{pytorchlightning}, torch geometric~\cite{torchgeometric} and the transformers library~\cite{wolf-etal-2020-transformers} for an efficient implementation of our proposed Twitter bot detection framework. We present our hyperparameter settings in Table \ref{implement} to facilitate reproduction.
Our implementation is trained on a Titan X GPU with 12GB memory. Our implementation is publicly available on GitHub.
\footnote{\url{https://github.com/BunsenFeng/BotHeterogeneity}}.

\subsection{Experiment Results}
We firstly evaluate whether these methods involve deep learning, leverage user interactions, learn user representation, involve graphs and graph neural networks or leverage Twitter heterogeneity. We then benchmark these bot detection models on TwiBot-20~\cite{feng2021twibot} and present results in Table~\ref{bigtable}. It is demonstrated that:
\begin{itemize}
    \item Our proposal consistently outperforms all baselines, including the state-of-the-art BotRGCN~\cite{feng2021botrgcn}.
    \item Successful graph-based methods, such as BotRGCN~\cite{feng2021botrgcn} and ours, generally outperform traditional approaches that do not consider the Twittersphere as graphs and networks. These results demonstrate the importance of modeling the topological structure of Twitter for bot detection.
    \item We propose the first heterogeneity-aware bot detection frameworks, which achieves the best performance on a comprehensive benchmark. These results bear out the necessity of leveraging Twitter heterogeneity and the effectiveness of our proposed approach.
\end{itemize}

In the following, we firstly study the role of graphs and heterogeneity in our proposed approach. We then examine the data efficiency and representation learning capability of our bot detection method.

\begin{figure}[t]
    \centering
    \includegraphics[width = 0.9\linewidth]{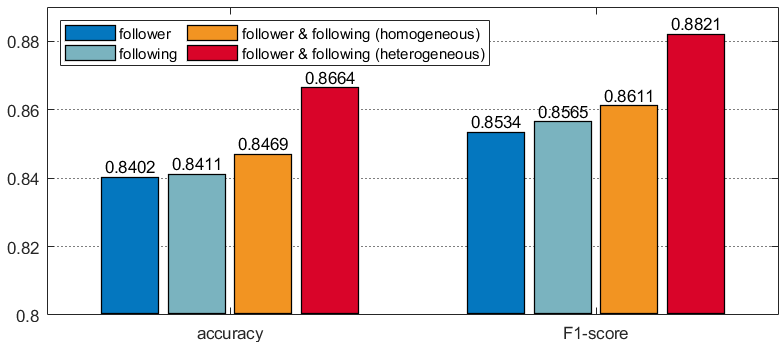}
    \caption{Ablation studying removing different parts of graph structure of our constructed Twitter HINs.}
    \label{edge_ablation}
\end{figure}

\subsection{Graph Learning Study}
We propose a graph-based bot detection model, which leverages the topological structure of the Twittersphere to capture subtle patterns and better identify bots. Specifically, we adopt user follower and following relationships as two types of edges that connects users as nodes to form a HIN. To prove the effectiveness of our proposed graph construction approach, we remove different types of edges and report results under these ablation settings in Figure \ref{edge_ablation}. It is illustrated that the complete graph structure, with both follower and following edges, outperforms any reduced settings. These results prove the effectiveness of our constructed HIN to model relation heterogeneity on Twitter.

\begin{table}[!t]
	\centering
	\caption{Ablation study of our proposed GNN architecture. RT and SA denote relational transformers and semantic attention networks respectively.}
	\begin{tabular}{c|c|c}
		\toprule[1.5pt]
		\textbf{Ablation Settings} & \textbf{Accuracy} & \textbf{F1-score}\\
		\midrule[0.75pt]
		full model & \textbf{0.8664} & \textbf{0.8821}\\
		\midrule[0.75pt]
		remove transformer in RT & 0.8521& 0.8679 \\
		remove gated residual in RT & 0.8478& 0.8646\\
		replace RT with GAT  & 0.8571& 0.8726\\
		replace RT with GCN & 0.8444& 0.8619 \\
		replace RT with SAGE & 0.8546 & 0.8687\\
		summation as SA  & 0.8512& 0.8654\\
		mean pooling as SA  & 0.8512& 0.8663\\
		max pooling as SA  & 0.8495& 0.8629\\
		min pooling as SA  & 0.8555& 0.8704\\
		\bottomrule[1.5pt]
	\end{tabular}
	\label{architecture_ablation}
\end{table}

Upon obtaining a HIN, we propose relational graph transformers to propagate node messages and learn representations. To prove the effectiveness of our proposed GNN architecture, we conduct ablation study on relational graph transformers and report results under different settings in Table \ref{architecture_ablation}. It is demonstrated that transformers, the gate mechanism and the semantic attention networks are all essential parts of our proposed GNN architecture.

To sum up, both our constructed HIN and our proposed GNN architecture contribute to our model's outstanding performance, which bears out the effectiveness of our graph-based approach.

\subsection{Heterogeneity Study}
\label{subsec:heterogeneity}
Our bot detection proposal models the intrinsic heterogeneity of Twitter to identify subtle anomalies of bots and conduct robust bot detetcion. We study the effects of incorporating heterogeneity and present our findings.

\subsubsection{Relation Heterogeneity}
\begin{figure}[t]
    \centering
    \includegraphics[width = 1\linewidth]{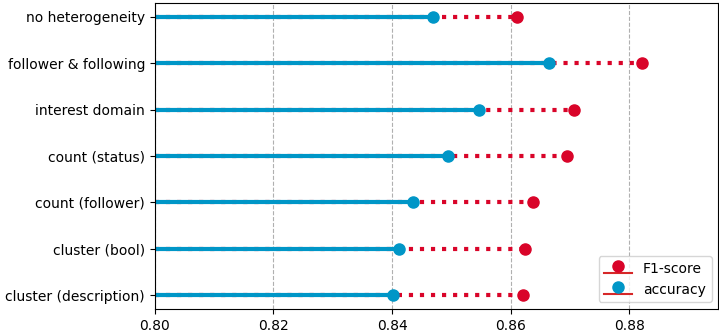}
    \caption{Performance of our proposed method with different relation heterogeneity settings. We cluster users with description and bool features, divide users with follower and status counts, leverage user interest domain in the dataset as well as following information to construct HINs.}
    \label{fig:relation1}
\end{figure}

Relation heterogeneity refers to the fact that there are diversified relations between users on the real-world Twittersphere. Our bot detection model incorporates relation heterogeneity by constructing HINs and leveraging them with relational GNNs. Different HINs could be constructed with different relation sets $R$, thus we propose different relation heterogeneity settings and present their results in Figure \ref{fig:relation1}. It is illustrated that most heterogeneous relation settings outperform their homogeneous counterpart, which proves the necessity of modeling relation heterogeneity for Twitter bot detection.

\begin{figure}[t]
    \centering
    \includegraphics[width=1\linewidth]{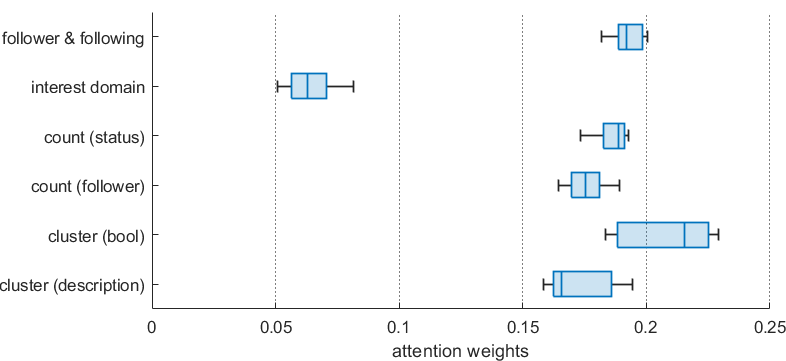}
    \caption{Attention weights of different sets of relations that co-exist on the real-world Twittersphere. We present the first, second and third quartile of results from multiple runs.}
    \label{attention_weights}
\end{figure}

To identify relation types that are crucial in Twitter bot detection, we combine all relations to form a comprehensive graph and use weights from the semantic attention networks to identify significant relations. Experiment results in Figure \ref{attention_weights} demonstrate that most heterogeneous relations contribute equally to our method's performance, while the user interest domain information in the data set is not as effective.

To sum up, we improve bot detection performance by incorporating relation heterogeneity and most relations are significant in our method's decision making.

\subsubsection{Influence Heterogeneity}
\begin{figure}[t]
    \centering
    \includegraphics[width = 1\linewidth]{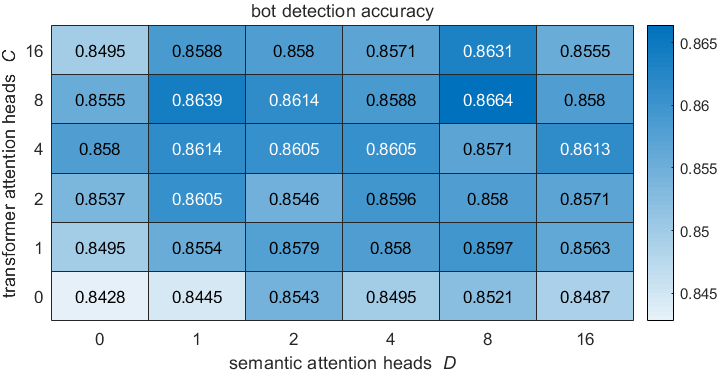}
    \caption{Ablation study of the attention mechanism in relational graph transformers and semantic attention networks.}
    \label{attention_heads}
\end{figure}

Influence heterogeneity refers to the fact that Twitter users have different patterns and intensity of influence over others on social media. We leverage influence heterogeneity with the multi-head attention mechanism in relational graph transformers. To validate the effectiveness of this approach, we conduct ablation study on the attention mechanism and present results in Figure \ref{attention_heads}. It is illustrated that incorporating the attention mechanism ($C > 0$, $D > 0$) outperforms methods without it ($C = 0$, $D = 0$). Besides, adopting multi-head attention networks ($C > 1$, $D > 1$) generally outperforms their single-head counterparts ($C = 1$, $D = 1$), proving the effectiveness of our design choices.

\begin{figure}[t]
    \centering
    \includegraphics[width=1\linewidth]{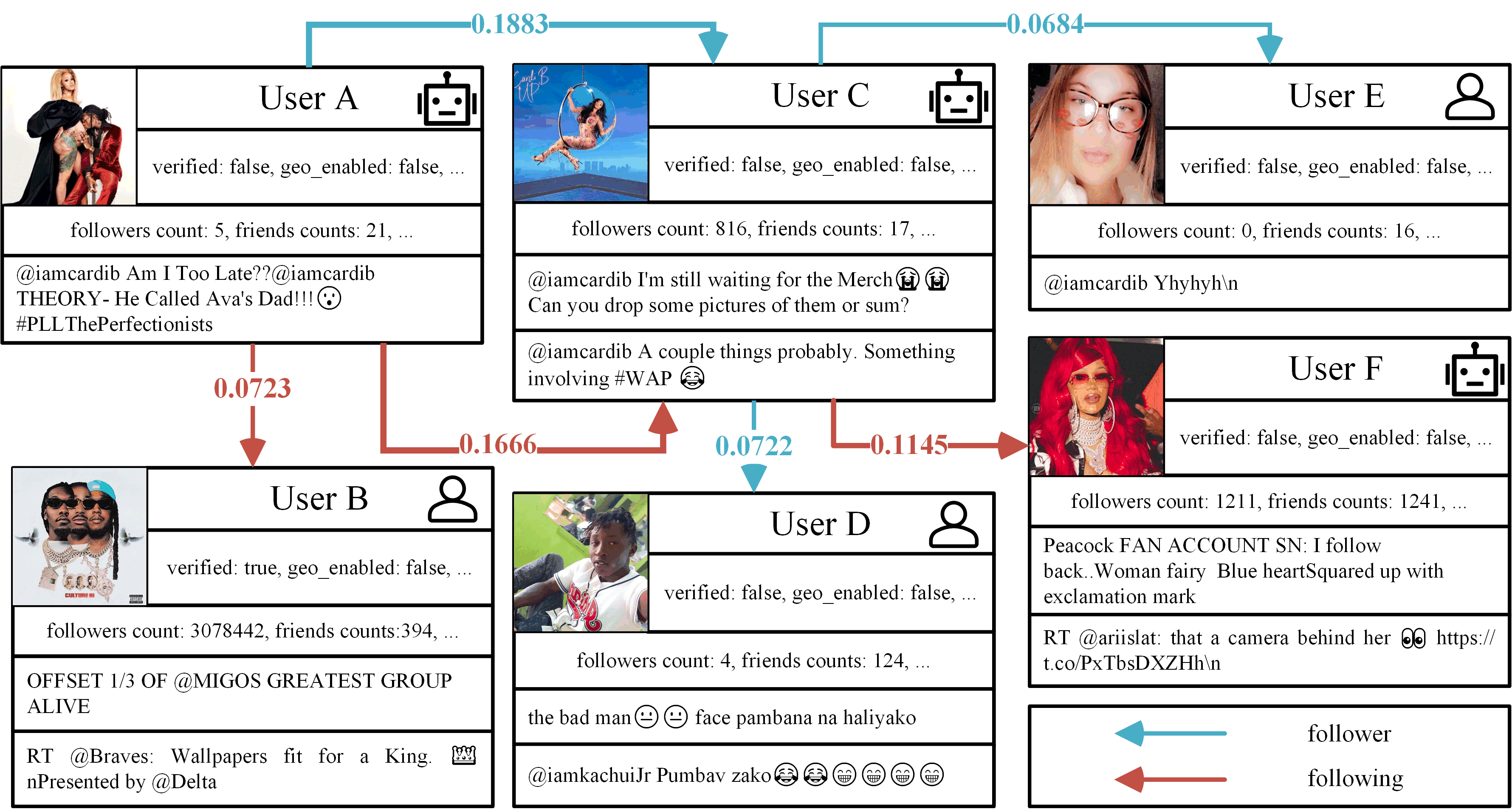}
    \caption{Example cluster of six real-world Twitter users and the attention weights between them.}
    \label{case_study}
\end{figure}

After proving the necessity of leveraging influence heterogeneity, we study a specific cluster of Twitter users and present their attention weights in Figure \ref{case_study}. It is illustrated that influence weights between bots are generally larger. By modeling influence heterogeneity, our method identify bots that act in groups and substantially influence each other.

To sum up, we improve bot detection performance by leveraging influence heterogeneity, while attention weights between users in the network yield valuable insights into our model's decision making.

\subsection{Data Efficiency Study}
\begin{figure}[!t]
    \centering
    \includegraphics[width = 1\linewidth]{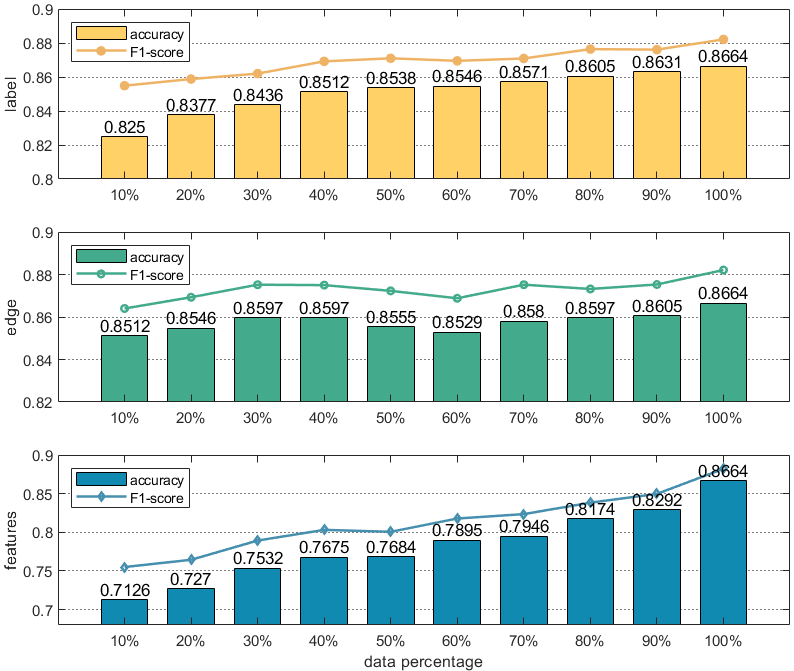}
    \caption{Model performance with limited data annotation, user interactions and user features.}
    \label{data_efficiency}
\end{figure}

Existing bot detection models are generally supervised and rely on large quantities of data annotations, while bot detection data sets are generally limited in size and labels. To examine the data efficiency of our bot detection model, we present performance with partial training sets, randomly removed edges and masked user features in Figure \ref{data_efficiency}. It is illustrated that our method would still outperform the state-the-of-art BotRGCN~\cite{feng2021botrgcn} with as little as 40\% training data and is also robust to changes in user interactions. Model performance drops significantly with reduced user features, which suggest that Twitter bot detection still rely on comprehensive analysis of user information in addition to the graph structure.

\subsection{Representation Learning Study}
\begin{figure}[t]
    \centering
    \includegraphics[width = 1\linewidth]{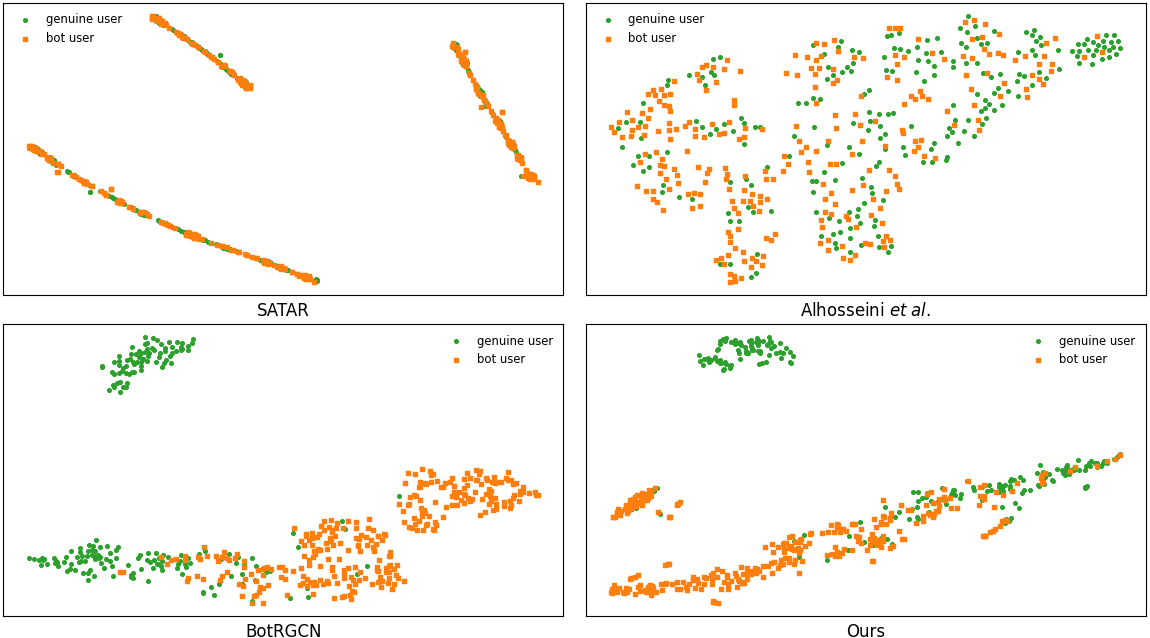}
    \caption{Plots of user representations learned with our method and different baselines.}
    \label{representitive}
\end{figure}

Our model, as well as few baselines, learn representation for Twitter users and identify bots with them. To examine the quality of representation learning with our proposal, we present the t-sne plot of user representation of our method and baselines in Figure \ref{representitive}. It is illustrated that our result shows higher levels of collocation for groups of genuine users and Twitter bots, which indicates that our method learns high-quality user representation.

\section{Conclusion and Future Work}
Twitter bot detection is an important and challenging task. We proposed a graph-based and heterogeneity-aware bot detection framework, which constructs HINs to represent the Twittersphere, adopt relational graph transformers and semantic attention networks for representation learning and bot detection. We conducted extensive experiments on a comprehensive benchmark, which demonstrates that our method consistently outperforms state-of-the-art baselines. Further exploration proves our method's graph learning strategy and the inclusion of Twitter heterogeneity are generally effective, while also performs well with limited data and learns high-quality representation for Twitter users. We plan to experiment with more diversified ways to model the Twittersphere as graphs and extend our graph-based bot detection approach in the future.

\section{Acknowledgments}
This work was supported by National Nature Science Foundation of China (No. 61872287, No. 62137002 and No. 62050194), CCF-AFSG Research Fund, Innovative Research Group of the National Natural Science Foundation of China (No. 61721002), Innovation Research Team of Ministry of Education (IRT\_17R86), Project of China Knowledge Center for Engineering Science and Technology and The Consulting Research Project of Chinese Academy of Engineering 
``The Online and Offline Mixed Educational Service System for The Belt and Road'' Training in MOOC China. We would also like to thank Zilong Chen for his help during the rebuttal, Herun Wan and Ningnan Wang for their contributions in preceding works, and all LUD lab members for our collaborative research environment.

\bibliography{aaai22}
\balance

\end{document}